# Social, Environmental, and Technical: Factors at Play in the Current Use and Future Design of Small-Group Captioning

EMMA J. McDONNELL, University of Washington, USA

PING LIU, University of Washington, USA

STEVEN M. GOODMAN, University of Washington, USA

RAJA KUSHALNAGAR, Gallaudet University, USA

JON E. FROEHLICH, University of Washington, USA

LEAH FINDLATER, University of Washington, USA

Real-time captioning is a critical accessibility tool for many d/Deaf and hard of hearing (DHH) people. While the vast majority of captioning work has focused on formal settings and technical innovations, in contrast, we investigate captioning for informal, interactive small-group conversations, which have a high degree of spontaneity and foster dynamic social interactions. This paper reports on semi-structured interviews and design probe activities we conducted with 15 DHH participants to understand their use of existing real-time captioning services and future design preferences for both in-person and remote small-group communication. We found that our participants' experiences of captioned small-group conversations are shaped by social, environmental, and technical considerations (e.g., interlocutors' pre-established relationships, the type of captioning displays available, and how far captions lag behind speech). When considering future captioning tools, participants were interested in greater feedback on non-speech elements of conversation (e.g., speaker identity, speech rate, volume) both for their personal use and to guide hearing interlocutors toward more accessible communication. We contribute a qualitative account of DHH people's real-time captioning experiences during small-group conversation and future design considerations to better support the groups being captioned, both in person and online.

This work is supported by the National Science Foundation, under grant IIS-1763199 and grant DGE-1762114

Author's addresses: Emma J. McDonnell, University of Washington, USA, ejm249@uw.edu; Ping Liu, University of Washington, USA, pingliu92@uw.edu; Steven M. Goodman, University of Washington, USA, smgoodmn@uw.edu; Raja Kushalnagar, Gallaudet University, USA, raja.kushalnagar@gallaudet.edu; Jon E. Froehlich, University of Washington, USA, jonf@cs.washington.edu; Leah Findlater, University of Washington, USA, leahkf@uw.edu

# 1 INTRODUCTION

Real-time captioning provides vital spoken conversation access for many d/Deaf and hard of hearing (DHH) people. Both human-generated and automatic captions have received substantial attention from HCI and CSCW researchers, with a focus on captioning in classrooms or other formal environments (e.g., [13,37,43]). While human transcription (e.g., CART) is most common in these settings, researchers have also examined automatic captioning, particularly for constrained environments like classroom lectures; commonly there is a single dominant speaker in these settings and the vocabulary used is more predictable [1,37,71]. In contrast, captioning for more informal small-group and one-on-one interactions has received less attention, despite the fact that human captioning's high cost ($60-$200 an hour in the US) and need for advance scheduling create significant barriers in this context [73,75].

While automatic captioning is an increasingly viable alternative to human captioning [38,61], its accuracy varies from 9-37% word error rates across tools [38]. Unlike human captioners, automated techniques cannot convey non-speech context (i.e., visual references, emotion, emphasis) nor can it intervene to improve communication. Moreover, small-group conversation's interactive nature, flexible social dynamics, and high level of spontaneity further limit existing captioning services. Ultimately, captioning and other access tools (e.g., paper and pen, texting, notes apps) all come with limitations and do not fully support DHH people during small-group conversations [20].

Despite the sociotechnical nature of small group captioning, most prior work has only examined *technical* considerations, such as how to convey uncertainty in automatic captioning through the use of simulated conversation in controlled experiments [7,59,60]. Seita et al. offer exceptions that explore how *social* interactions and behaviors impact captioning [65–67]. They first found that hearing people speak more loudly, clearly, quickly and with non-standard articulation when they are being captioned in small-group conversations [65]. In a preliminary [67] and follow up study [66], Seita et al. had a hearing actor modulate their conversation behaviors in various ways as part of a controlled experiment (e.g., speech rate, voice intensity, eye contact), and measured what behavior variants DHH participants preferred (e.g., fast, medium, or slow speech), and which behaviors were most important. They provide quantitative evidence that hearing people's behaviors impact DHH people's experiences of one-on-one captioned and interpreted conversations. These findings motivate the need to more deeply understand DHH people's small group captioning experiences through a sociotechnical lens. In this paper we address the questions: *What social, environmental, and technical factors impact the use and usefulness of captioning in small groups? What opportunities exist to design captions and caption displays in ways that support more accessible group communication practices?*

To begin addressing these questions and to ground future small group captioning technologies in the needs and desires of DHH people, we conducted an interview and design probe study with 15 DHH participants. Each session began with an interview covering the participant's experiences with real-time captioning in small-group conversation and their perspectives on the role of hearing conversation partners in creating or obstructing accessibility. Participants then completed a design probe activity, building on methods outlined in [24,26,33,51]. In this activity, we presented a series of potential future captioning features (e.g., displaying speech rate, flagging overlapping speakers, supporting error correction by hearing people) to provoke discussion around what new designs participants desire and how that technology could be integrated into small-group conversations.

Our findings highlight the myriad social (e.g., group norms, preferred communication modes), environmental (e.g., furniture configuration, online availability of a text chat), and technical (e.g., caption lag, built-in speaker identification) factors that shape real-time captioning, contributing an understanding of the context that surrounds captioned conversation. Particularly, we find that: (1) captioning's efficacy is highly determined by the group being captioned, (2) current captioning tools are often insufficient during interactive conversation, and (3) while the lack of visual and spatial information online create barriers, features of video conferencing also provide new opportunities to increase access. Participants' responses to the design probe activity also highlight the potential to create more captioning-friendly environments, both online and in person, and suggest that providing conversation feedback and warnings to guide captioning-friendly group norms is a promising direction for future development. Based on these findings, we discuss the need to consider the intersection of social, environmental, and technical factors in captioning research, propose a reframing of captioning as a group technology, and put forth future design guidelines that center DHH peoples' needs.

More broadly, we contribute (1) an empirical account of DHH participants' experiences of small group captioning which highlights how social, environmental, and technical factors impact its use and efficacy, (2) an exploration of design opportunities to support small group captioned conversations and future design guidelines, (3) an understanding for both (1) and (2) of how online environments—a historically little-studied captioning context—shape captioning experiences and preferences, and (4) reflections on reframing captioning as a group technology.

## 2 RELATED WORK

To contextualize our study, we analyze current captioning methods, caption use and design, and provide a Deaf and disability studies framing.

### 2.1 Real-Time Captioning Services

DHH people use a variety of real-time captioning technologies, each with their own tradeoffs. CART—human-generated verbatim captioning—is the most popular and claims to be at least 98% accurate for all words typed [74], but this includes after-the-fact corrections; the accuracy of live CART is lower [37]. Moreover, CART is expensive and must be scheduled in advance [73]. An alternative, C-Print, summarizes content within sentences and uses a shorthand style [76], but is also costly (~$60/hour) and must be pre-scheduled [19]. Both CART and C-Print are frequently provided by in-person and remote transcriptionists. Crowdsourcing has been explored to allow non-experts to generate high-quality captions [29,46], but even the most developed system, Legion:Scribe, remains in private beta release [77].

While human transcription remains the legally protected standard for captioning in the US and around the world [55,75], automatic captioning using automatic speech recognition (ASR) is increasingly used for informal interactions and when accommodations are not otherwise available [30]. Tools such as Otter.ai and Google's Live Transcribe provide free or low-cost captioning but accuracy can be a concern: recent analyses found that Google's API outperforms other ASR, recording average word error rates around 9% [22,38]. Additionally, ASR performance deteriorates in complex audio environments [69] and does not handle accents well, including Deaf accents—Glasser et al. [25] found that Microsoft's Translator Speech API's word error rate was 18% for hearing speakers and 78% for Deaf speakers. Unlike human transcription, ASR does not convey non-speech information such as laughter or consider high-level context, such as a child trying to say a new word. Furthermore, many within the Deaf community oppose using automatic captioning in place of human transcription, considering it to be insufficient access [16]. These services *generate* captions, but how people use captions and how to design effective captioning displays are additional research questions—and the focus of our study.

### 2.2 Caption Design and Use

We review work on DHH people's experiences with captions, focusing on caption design and interactions between DHH and hearing people.

Captions are an imperfect technology and prior work has documented challenges and potential solutions. A key concern when designing captioning systems is limiting visual attention split, and researchers have explored myriad display configurations to enhance DHH people's ability to read captions while attending to other aspects of conversation, including integrating captions into the environment, using head-mounted displays, and annotating captions [15,33,34,42,47,52,57,59]. Additional design efforts seek to account for the fact that captioning flattens expressive elements of speech: adding punctuation to automatically generated captions can improve readability [28,70], placing captions near speakers in videos and displaying speech volume has been well-received by DHH viewers [31], and both research prototypes and commercial tools use color to differentiate captioned speakers [27,78].

There are also known hurdles to caption comprehension that remain unaddressed. Jensema [35] identified that the ideal captioning speed is around 145 words per minute (wpm), with a drop off in comprehension after 170 wpm (typical human English speech rates are 120-160 wpm [62]). Additionally, a key concern around ASR-generated captioning is high error rates, and many researchers have explored ways to communicate this uncertainty to caption viewers [5–8,59,60,65,68]. We use the literature above to inform our design probes and to recommend designs that could address the challenges expressed by our participants.

To improve captioning design, HCI work has often introduced new technologies to classroom settings (e.g., [1,13,15,19,21,39–43,71]). This classroom context differs from small-group conversations, which in contrast tend to be less structured, have multiple speakers rather than a primary lecturer, and are often not well-supported via formal accommodations (e.g., CART, interpreting). Several studies have explored small group captioning needs via simulated one-on-one conversations [6,8,59,60] and while they provide insight into caption preferences (e.g., use 2 lines and common fonts [6]), their non-interactive nature does not allow for understanding how captions influence small-group social dynamics. Other research has explored the viability of phone-based ASR combined with typed responses, having Deaf and hearing participants communicate in the lab [18] or field [49], and their findings, though brief, have been positive. Further, some head-mounted displays for captioning have been evaluated in small-group conversation, showing that participants benefit from seeing captions in the same field of view as their conversation partner(s) [34,57].

Compared to this existing body of captioning research, our interview and design probes treat captioning as a technology used by groups, which opens new questions about conversation participants' impact on captioning success and the potential for caption designs to shape individual and group behaviors. Accessibility research in other contexts has begun to explore such an approach, e.g., to help ASL interpreters and classroom instructors coordinate content [9] and to assist presenters in increasing non-visual accessibility [58]. Additionally, captioning tools to date have been predominantly studied in the context of in-person conversation, with the exception of Kushalnagar and Vogler's teleconference best practices [44], while our study examines captioning both in person and with online video calls.

As mentioned in the Introduction, most relevant to our paper is work on understanding and designing to support interaction between DHH and hearing people during caption use. Seita et al. [65] studied automatically captioned small-group conversation between DHH and hearing people, finding that, in the presence of captions, hearing people altered speech characteristics, such as volume and rate, but the study did not report on DHH participants' experiences or the social impacts of using captioning, areas our research explores. Seita and Huenerfauth [67] also conducted a controlled experiment with 8 DHH participants to explore the impact of a hearing researcher modulating their speech in several ways (i.e., speech rate, volume, eye contact). At least some participants noticed each of the six behaviors, with open-ended comments suggesting that speech rate is particularly important but that all behaviors were relevant. In a 2021 follow up study [66], Seita et al. repeated their methodology from [67] with 20 DHH participants in person, finding that modulations in intonation and enunciation statistically significantly impacted participants' satisfaction with the hearing person's behavior and that
enunciation and eye contact where more important than intermittent pausing. They also ran
this study with DHH participants online, using ASL interpretation rather than captioning to
convey the hearing actor's meaning. Combined, these studies help to motivate our
sociotechnical analysis of captioning use by beginning to show that hearing people adapt their
speech in the presence of captions and that DHH people notice some of these adaptations.
Building on that work, we conduct in-depth interviews on DHH people's small group
captioning experiences and explore potential designs to guide group communication behaviors.

Finally, we draw on qualitative work that contextualizes caption use. Kawas et al.'s [37]
analysis of real-time captioning in the classroom identifies that most hurdles students face are
fundamentally sociotechnical, requiring technological, social, environmental, and policy
solutions, and we are inspired to explore small group captioning with a similar sensibility.
Complementing our work is a study from Wang and Piper [72], who interviewed and observed
existing dyads of Deaf and hearing collaborators, focusing on interactions when
accommodations are unavailable (i.e., not focused on captions). They found that, over time,
these Deaf-hearing teams create accessible practices, including flexibly switching between
spoken and written language, learning to prioritize shared visuals, and providing ad hoc,
informal transcription and sign language interpretation. We explore their theory of accessibility
as a co-created group practice among Deaf-hearing teams in the context of captioned
conversations and with DHH participants who use a wider variety of communication styles
(their DHH participants all both signed and voiced).

## 2.3 Deaf and Disability Studies Perspectives

While the hearing world often thinks of deafness as an audiological diagnosis, many people
identify as capital-D Deaf, signaling engagement with the Deaf community and Deaf culture.
Deaf studies scholar Harlan Lane argues that Deaf identity is akin not to a disability but to an
ethnicity, with its own linguistic, conversational, and cultural norms [45]. Deaf studies names
audism as systemic discrimination on the basis of hearing ability, identifying the structural
barriers that DHH people face as the fault of oppression from hearing people and institutions
[2]. Because accessible technology developments for DHH people can all too easily perpetuate
audism [23], we ground our research in Deaf studies critique.

While respecting the contested cultural differences between deafness and disability, our
approach to caption design is also impacted by disability studies thinking on accommodations.
A key contribution of disability studies is the notion of models of disability, commonly
contrasting the medical model, conceptualizing people as intrinsically disabled by abnormal
bodies that need fixing, with the social model, framing disability as what happens when an
ableist society does not meet the needs of people with impairments [56]. Mankoff et al. [50]
highlight how moving away from the medical model in assistive technology design better
supports disabled people. Following this model shift, Kasnitz [36] argues for *"community-based
accommodation."* This reconceptualization treats hearing and Deaf people in conversation as
equally reliant on accommodations and reimagines who is responsible for arranging access. We
also turn to interdependence, the move from treating disabled people as fundamentally
dependent to viewing all people as inter-reliant [53], which Bennett et al. [4] argue can lead to
assistive technology that treats access as relational, challenges ability hierarchies, and highlights
disabled people's competencies. Our study is guided by this body of work to focus on the social
experience of captioning and to consider designs of future captioning systems that design for
the group, not just DHH individuals.

| ID | Age | Gender | Identity | Preferred Method of Communication | Frequency of captioning use | Frequency of oral communication |
|----|-----|--------|----------|-----------------------------------|------------------------------|----------------------------------|
| P1 | 54 | M | deaf, having hearing loss | Oral, written | A few times a week | All the time |
| P2 | 26 | F | deaf, hard of hearing | Sign, oral, written | A few times a week | Most of the time |
| P3 | 44 | F | Deaf | Sign | 2-3 times a month | Never |
| P4 | 34 | F | Deaf | Sign, written | Multiple times a day | Never |
| P5 | 71 | F | deaf | Written | Multiple times a day | Some of the time |
| P6 | 24 | F | Deaf | Sign | Multiple times a day | Some of the time |
| P7 | 47 | M | Deaf | Sign | Multiple times a day | Most of the time |
| P8 | 30 | F | Deaf | Sign | Multiple times a day | Some of the time |
| P9 | 69 | M | Deaf | Oral | About once a day | Most of the time |
| P10 | 53 | F | Deaf | Sign, written, texting | About once a day | Never |
| P11 | 70 | F | Hard of hearing | Oral, written | Multiple times a day | All the time |
| P12 | 21 | NB | deaf | Oral | Multiple times a day | All the time |
| P13 | 56 | F | Deaf | Sign | About once a day | Some of the time |
| P14 | 28 | M | deaf | Sign | Multiple times a day | Infrequently |
| P15 | -- | F | Deaf | Oral | 2-3 times a month | All the time |

Table 1. Summary of participant demographics, as reported in the pre-study session survey. P15 chose not to disclose her age.

## 3. METHOD

To understand DHH people's experiences using captions in small group scenarios and their preferences for future captioning systems, we conducted individual qualitative study sessions with 15 DHH participants. We intentionally recruited only DHH participants for this research because our study design is shaped by a commitment to placing the power to shape design recommendations for future captioning tools in the experiences, desires, and needs of the DHH community. The study was conducted remotely via videoconferencing and had three components: a pre-session survey, a semi-structured interview, and a design activity.

### 3.1 Participants

Participants were recruited via email lists at two US universities, social media, and snowball sampling. We required that participants be 18 years or older, able to participate in a Zoom call, self-identify as d/Deaf or hard of hearing, and frequently use real-time captioning—either automated or via a human transcriptionist—for conversation access. We recruited 15 DHH participants (4 men, 10 women, 1 non-binary person), a sample size in line with community norms for similar studies and appropriate for reflexive thematic analysis [12,14,48]. On average, participants were 44.8 years old (*SD*=17.9, *range*=21-71)—see Table 1. Participants had a wide range of *"preferred communication methods"*: sign language (60.0%), oral (40.0%), and written (46.7%) communication (participants could select multiple communication preferences). They also had differing experiences with spoken conversations, with participants communicating orally all the time (26.7%), most of the time (20.0%), some of the time (26.7%), infrequently (6.7%), and never (20.0%). Frequency of captioning use ranged from multiple times a day (46.7%) to 2-3 times a month (13.3%).

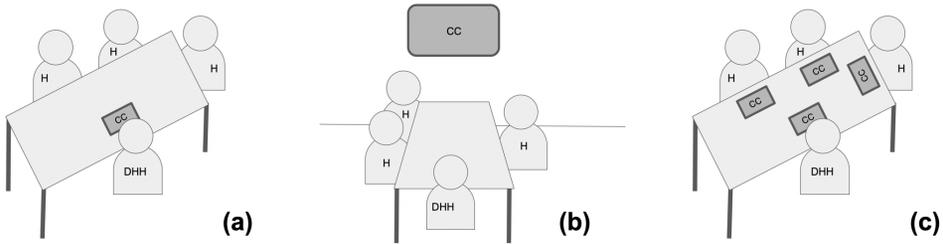

Figure 1: Participants discussed captioning displayed for only DHH people via personal devices (a), or for
all meeting attendees via a large screen/projector (b) or personal devices (c).

### 3.2 Procedure

All interviews were conducted remotely due to the COVID-19 pandemic during the summer of 2020. Prior to meeting with researchers, participants completed a ~20-minute online survey providing demographic information, background on their experience with captioning, context on how they access captioning services, and their perspectives on the technical, environmental, and social factors that impact their experiences using captioning. The study session took 90 minutes, beginning with a semi-structured interview (~35 minutes) followed by a set of design activities (~55 minutes). All study sessions were conducted by the hearing first author and facilitated via participants' preferred accommodations: eight participants chose ASL interpreters, five chose CART, one chose automatic captioning, and one chose neither interpreting nor captioning. The researcher screen-shared a slide deck with the study instructions, questions, and design probes, both to be able to discuss design ideas remotely and to allow for multiple ways to access study materials. See Supplementary Materials for the full slide deck.

The semi-structured interview focused on how participants use captioning in their daily lives and how social factors shape their experiences. Questions covered experiences with different captioning services (e.g., CART, automatic captioning), when captioning works well or poorly, when captions are unavailable but would be helpful, how hearing people help or hinder captioning, and how comfortable they are asking hearing people to adopt new communication practices. Throughout, the researcher asked participants to reflect on captioning use both in person and remotely.

Following the interview, the researcher facilitated a design probe activity with each participant, inspired by the use of this method in other papers, including accessibility work with DHH participants [24,26,33,51]. Design probe investigations afford light-weight investigation of future technologies and allow researchers to get participant input before committing to a specific design [32], making this method well-suited to our research questions.

The activity included three sets of probes, which were designed to act as a starting point for discussion about future captioning setups, including ideating on potential new features and caption correction systems to be used during small-group captioned conversation. Specifically, we introduced the design probe activity by asking participants *"to try to envision captioning in the future"* and clarifying that *"we don't have to be limited to how technology currently works."*

Throughout, we grounded the discussion in the context of being the sole DHH person *"using automatic captioning during an in-person meeting with a small group of hearing people."* We also asked participants to contrast their in-person responses to online contexts. This activity included probes exploring the following:

1. *Caption visibility.* The first probe centered on participants' preferred method for viewing captions. To introduce this probe, we asked participants to describe their ideal captioning set up for the scenario, prompted with *"for example, you might think about the room setup, where the captions are displayed, who sees them, and anything else."* We then showed captioning setups that varied in who could see them: captioning on the personal device of only the DHH person (Figure 1a), captioning on a large screen/projector that all meeting attendees could see (1b), and captioning on the personal devices of all meeting attendees (1c). We had participants consider the advantages and disadvantages of each setup, and how they would feel about analogous setups for remote meetings.

2. *Additional features.* The second probe focused on adding information to the captioning display. Following a general introduction to this focus, we described five potential display features: speech rate, speaker identity, volume of a speaker's voice and of background noise, caption lag, and a multiple concurrent speaker warning. We selected each probe based on current captioning practices as of June 2020, prior work, and knowledge from our team of Deaf and hearing researchers: human captioners often convey speaker identity and overlap, our team identified lag as a significant consideration during small-group captioned conversation, and prior work has identified speech rate [35,67] and volume [31,67] as of interest. For each feature, the researcher first introduced the idea (e.g., *"Speech Rate: Show how fast the speaker (you or others) is talking"*) and asked participants what they thought about showing this information in some way during in-person meetings. To make the idea more concrete, the researcher then showed a specific design mockup (Figure 2a-e), which we described as a *"rough example of how this feedback could look,"* and asked participants: (1) to share any other ideas they had about how the information could be shown, (2) who (if anyone) they would want to see that information, and (3) how they would feel about this type of information being included in online vs. in-person meetings. After viewing all five sound qualities, participants ranked them and had the opportunity to suggest other information. Note that when creating the design mockups, we opted to display information directly rather than via abstraction (e.g., showing caption lag in seconds delayed rather than as a warning to wait for captions to catch up), to act only as a starting point for discussing how the information could ultimately be displayed.

3. *Caption corrections.* The third probe focused on allowing meeting attendees to correct captioning errors in real time. To elicit conversation around this idea, we first introduced the concept and discussed it in the abstract before showing a mockup of a system where meeting attendees could type corrections for errors they notice (Figure 2f). We intentionally kept the specifics of this mockup vague to explore how participants would imagine such a system could work. We asked participants for feedback on the idea for both in-person and online meetings.

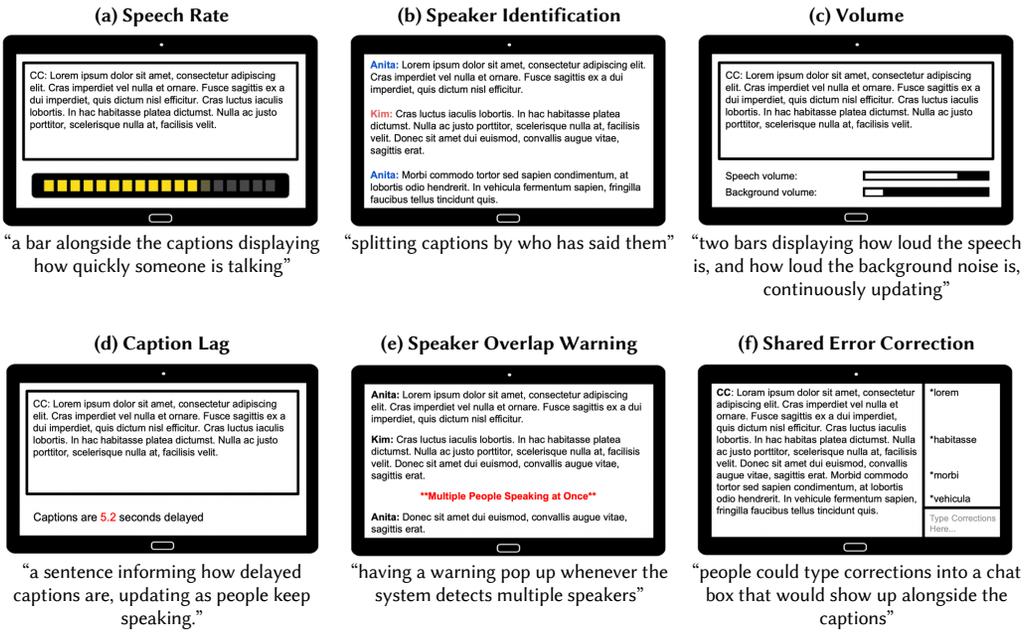

Figure 2: As a design probe, participants were shown mockups that we described as "a rough idea of how" each potential feature could be implemented: (a) speech rate, (b) speaker identification, (c) volume, (d) caption lag, (e) speaker overlap warning, and (f) shared error correction.

Finally, participants sketched out their ideal captioning setup for both in-person and online meetings using a pen and paper. Participants shared their sketches by holding them up to the camera, describing them, and sharing ideas they could not capture in the drawing itself. To close the session, participants were given the opportunity to ask any questions, were compensated with a $50 Amazon gift card, and were asked to email their sketches to the researcher.

### 3.3 Analysis and Positionality

All interview data was transcribed, either directly using the CART transcripts for sessions where a human captioner was able to caption both the researcher and participant (*N*=4/15) or post hoc by either the first author or a transcription service. We analyzed the transcripts using reflexive thematic analysis as outlined by Braun and Clarke [10,11] in combination with a summary of participants' reactions to new designs, which helps identify cross-cutting themes as well as synthesize concrete design recommendations. Our thematic analysis is semantic and realist, with a mixed inductive and deductive approach to the data. Thematic analysis emphasizes that findings are not waiting to be discovered but are actively shaped by the research team and their own biases. The first author, who ran all interviews and led analysis, is hearing and an ASL beginner. Some authors, who were involved in study design, analysis, and writing, are Deaf.

### 4. FINDINGS

Guided by our research questions, we (1) highlight themes we identified in participants' experiences of small group captioning, emphasizing the influence of technical, environmental, and social factors, and (2) report on participants' reaction to our design probes, providing design considerations around better supporting DHH captioning users and engaging their hearing interlocutors in making small-group conversation more accessible. As the study took place during the COVID-19 pandemic, we report on experiences using captioning for online and in-person small-group conversation.

## 4.1 Current Experiences of Captioning

To understand how intersecting social, technical, and environmental factors shape DHH people's use of captioning, our analysis highlights: (1) the role of interlocutors in making conversations (in)accessible, (2) mismatches between the capacity of current technology and the demands of *interactive* small-group discussion, and (3) specific considerations for captioning online conversation.

### 4.1.1 Social Impacts on Captioning

Our participants described their experiences of captioning as highly determined by *social* dynamics: some groups develop collective, adaptive norms around captioning, while others may limit conversation access through unwittingly inaccessible behavior or explicit judgement of DHH captioning users.

**Adaptive group practices.** Both during informal interactions with automatic captioning and formal meetings with CART, participants described the benefits of hearing collaborators' willingness to alter communication styles. Participants who prefer not to voice (8 of 15 participants) explained that to use captions during interactive conversation they write out their contributions while the person they are communicating with is captioned, often using speech-to-text on the DHH person's phone. While transitioning between typing and captioning has a different rhythm than most spoken conversation, P10 found that hearing interlocutors *"get more to the point rather than wandering [in their] speech. I think that's an advantage."*

Participants who voice also benefited from interlocutors who were willing to change norms during spoken conversation. P1 found that his manager, who is *"extremely sensitive"* to the access gaps that persist while using CART, has made work meetings more accessible by taking advantage of online environments to correct captions in the meeting chat and intentionally pausing between topics so that he has a chance to jump in, socially adjusting for the technical limitations of captioning. P2's workplace has even more extensive group practices—to avoid interrupting speakers, they use a set of hand gestures to communicate when to slow down, speak up, or spell out uncommon words. P2 explains that her colleagues do so because having *"captioning available [is] not always going to be enough for someone. Your culture needs to change in order for the captioner to be more effective."* When all parties in a conversation are willing to adopt new social norms around communication, they create a distinct, more accessible solution.

**Unsupportive communicators.** Participants further highlighted the importance of behavior when describing interlocutors who actively or unwittingly made conversations less accessible. Several participants mentioned disengaging from conversations when hearing people speak over one another or speak too quickly to be captioned. For example, during P12's discussion-based classes, *"there have been a couple times where it's just been like, I don't understand this conversation so I'm just going to go home and wait for the transcript."* Participants also described how moments of acute judgement from others altered how they felt about using captioning tools thereafter. For instance, P11 had to verbally communicate with her notetaker during a group meeting and *"somebody stood up once and said, 'why are they in the corner talking?' I had to say because I can't hear. And it was just like, why was that even necessary? So after that, I just kind of wanted to do my own thing."* While P11's colleague may have been oblivious to the ramifications of their comment, P15 described the impact of active judgement and rigidity around conversation norms. She now joins work meetings via text relay, despite preferring to voice and having tendonitis in her hands, because *"my manager doesn't like to hear my voice. She told me the coworkers said I talk loud. [...] It makes me feel insecure."*

Though participants described benefits when hearing people figured out how to communicate accessibly, others explained that sometimes hearing people's instincts, such as slowing down, speaking loudly, or overenunciating, are not effective. For example, P12 commented that these adaptations can be *"done with good intentions but that's not always helpful [because] it makes me feel like you're not treating me as an equal sometimes."* P4 explains that in her experience, hearing people *"do care, it's just that they don't necessarily think about deaf people,"* which means that the burden, or *"constant scourge,"* of creating a captioning-friendly environment typically falls on her.

### 4.1.2 Technical Considerations While Captioning Interactive Conversation

Regardless of social support, captioning experiences differ with *interactive* conversation as compared to one-way communication (e.g., a lecture or seminar). Participants describe how technical aspects of captioning are ill-suited to the particular social dynamics of interactive conversation and the ways they use captioning as one of many access strategies.

**Technical mismatches in interactive contexts.** Our participants outlined aspects of interactive conversation that are not well-supported by real-time captioning. Participating in small-group interactive conversations requires being able to jump in during brief pauses and P1 explains that the delay inherent to captioning makes this difficult: *"everyone's still talking according to the screen, but they have finished. There's probably like an eight second lag. And sometimes I'm really anxious to say something or correct somebody and then I find that I'm interrupting somebody."* On top of temporal mismatches, captioning does not capture the speaker's tone, which P9 considered invaluable to avoid interrupting: *"I'm assuming some intonation [but] there is nothing on their face that indicates that they are going to complete their sentence."* Furthermore, while P2 found she could engage in captioned conversation with a small number of people, *"if the group gets bigger and other people are talking at the same time it's really hard to follow a conversation and it's also just as hard for the captioner."*

For participants who preferred not to voice, captioning alone does not adequately support interactive conversation. While captions worked when P14 did not need to reply, he explained that *"when I'm trying to say something, captioning doesn't really function for me in that capacity at all."* However, as automatic captioning has become more widespread, participants who would otherwise use interpreters reported experimenting with the technology. When P4's workplace suddenly shifted online due to COVID, delays with remote sign language interpreting services caused her to join meetings using automatic captions and text chat. P4 explained that, while not ideal, she was *"happy [she'd] found more than one solution,"* one that was enabled by the online environment, the technical capacity to turn on automatic captions, and social expectations that her typed contributions would be integrated into work meetings.

**Concurrent access strategies.** To manage the limitations of captioning for interactive conversation, participants often used other communication strategies in tandem. Some participants could mostly follow a conversation using their speechreading skills, residual hearing, and assistive listening devices, and they described using captions to augment their understanding, rather than as the primary way of accessing a conversation. For instance, P12 used Google's Live Transcribe when conducting interviews for a class project: *"if I couldn't understand what the other person was saying, [...] I would ask them to repeat it first. If I still didn't understand, I would just look down [at the app]."* Three participants described their use of captioning to augment sign language, such as P8's experience in discussion classes where *"some students had really lousy signings. So, I was able to look at CART instead."* The preference for flexible access strategies was shared by P15, who gets frustrated by CART writers who make her look at their captions, stating *"I have a right to lipread or look at the screen. It's my choice."*

### 4.1.3 The Environmental Affordances of Online Captioning

The environmental shift from in-person communication to online video conferencing introduces new social norms, a unique set of possible interactions (e.g., text chat), and different design and technical needs, all of which shape the experience of DHH captioning users. As this study was conducted in summer 2020, participants reflected on the sudden shift to online communication driven by the COVID-19 pandemic.

**Spatial and environmental considerations.** Many DHH people rely heavily on visual and spatial cues to follow and participate in conversation, and participants described challenges and gains that came with moving to online, two-dimensional space. P3 missed being able to spatially connect captions to a speaker like she would in person, while P13 lost out on being able to follow the gaze of *"the captioner [who] is going to probably look towards the person that's speaking."* However, P9 explained that while he struggled to match captions to the speaker on most platforms, Google Meet's speaker identification was *"fingers and eyes above everybody else that's doing captions."* Other participants echoed P2's experience, that online *"it's actually a lot easier to identify the speaker and easier to capture whoever is speaking at a time,"* and P4 and P11 mentioned the benefit of features that highlight the active speaker in online meetings.

**Toward inclusive conversation access.** Overall, participants found that moving life online brought with it features (including speaker identification) that have provided greater access. P1 stated that *"now that we can't go into the office it's been much more of an equalizing factor"* because his hearing colleagues are more motivated to limit overlapping speech and are also juggling lagging and malfunctioning technology. New online interaction paradigms also served to equalize conversation; for example, P8 discovered that automatic captions and an active text chat made it so that *"lots of people in the audience don't realize that I'm Deaf because we're all running on the same system at that point."* Several participants said automatic captions allowed them to access online meetings or social gatherings that would have otherwise been difficult to join in person. Furthermore, having text chat available at all times has created new opportunities: many participants' hearing conversation partners used the chat to correct mis-captioned jargon, P3 was able to use Microsoft Teams' messaging features to clarify confusing captions mid-meeting, and P2's friends used private chat to provide a transcript for her during uncaptioned Bible studies. These emergent social practices are enabled by the unique affordances of online environments.

### 4.1.4 Summary and Implications

Our findings show that DHH participants' experiences of captioned conversations are deeply shaped by social, environmental, and technical context. Participants' accounts of the impact of their hearing interlocutors demonstrate that captioning is a highly social technology and that the people being captioned are key stakeholders in determining conversational accessibility. These findings affirm Seita et al.'s focus on the interplay between DHH and hearing people [65–67] but suggest that relational contexts, which may not be captured in controlled lab settings, are crucial to negotiating accessibility when using captions. Wang and Piper [72] outlined how Deaf and hearing dyads adapt when communicating without accommodations and we find that collective adaptation remains critical even after captions are turned on. Additionally, the challenges participants described during interactive conversation show that captioning alone does not guarantee access, particularly for DHH people who do not voice. While some of these hurdles could be lessened with better technology, they are also fundamentally social. There is a growing body of work on one-on-one automatically captioned interactions in which DHH people type their contributions [e.g., 17,19,45,60] and future work could further explore emergent social norms during these interactions. Further, understanding that captioning is used in parallel with other access strategies for interactive conversation prompts consideration of how future captioning displays could better match their contexts of use. Finally, our participants' experiences suggest that online captioned conversations are occurring in an environment with fundamentally different affordances than in-person conversation. Some of these affordances, such as missing spatial information, pose new access barriers which designers have begun to address [44]. Yet many other aspects of online communication may be well-suited to captioned conversation and features, such as text chat, have been little explored but hold great potential for future accessibility.

## 4.2 Design Probe Findings

While the previous section reported on participants' experiences using captioning in small groups, here we turn to ideas and responses that arose during the design probe activity. As described in the Method section, these probes were meant to prompt participants to envision a range of possible future captioning designs. The probes were described simply as *"a rough example"* of how a particular idea could be instantiated in a captioning setup and were shown only after an initial conversation about each feature. We quantify positive/negative reactions as well as provide qualitative summaries.

### 4.2.1 Caption visibility

We asked participants to reflect on their ideal captioning setup and to consider the advantages and disadvantages of the following three in-person captioning setups and their digital analogs (Figure 1): captions available (1) only on the DHH person's device, (2) projected on a shared screen, and (3) available on the personal devices of all conversation participants. We posed these specific probes to gain a deeper understanding of how the type of display shapes captioning environments, to assess participants' feelings around making captions visible to their interlocutors, and to better understand how physical environments impact social dynamics.

**Personal device only.** Participants had mixed reactions toward having captions available to only themselves: some valued the autonomy and privacy of this setup, while others disliked it, describing feeling ostracized. For example, when considering how the display would impact conversation, three participants felt a personal display would be minimally disruptive because *"one person is usually pretty good at flicking between looking down and looking up"* (P12), but four others disagreed, arguing *"the hearing people would be able to see each other [...] but the deaf person is glued to the screen"* (P9). Four participants took issue with the assumptions built into personal displays, arguing these assumptions suggest that *"the deaf or hard of hearing person is the problem that needs to be fixed"* (P10).

**Shared caption display.** Participants largely saw value in a prominent, shared caption display but some worried that it would reshape the conversation environment in a way that negatively impacts social dynamics. Several participants (N=6) explained that setting up a shared display felt like an effort to equalize the conversation: *"Rather than remaining in that dominant space where they normally do, everyone is a little bit more aware of what life can look like for us"* (P8). Participants further identified benefits of shared captions: four referenced past experiences having their hearing interlocutors notice and correct caption inaccuracies and three considered that their hearing conversation partners may also want captions, especially those who are learning English or have audio processing disabilities. Others, however, had concerns, including difficulty managing captions and presentation slides (P1) and a loss of eye contact with the speaker (N=3): P13 explained that needing to look up at the screen means she misses *"the human connection part, that's important."*

**Captions on all personal devices.** Many of the advantages of a shared group display also applied to the third setup, displaying captions on everyone's personal devices. Almost half (N=7) of the participants saw distributed captioning as an equalizing force, with P3 favoring it because everyone has the *"same thing going on and it helps hearing people feel like they're part of the deaf individual's team."* Participants also saw technical benefits of this setup, such as allowing customization (P2) and potentially leveraging everyone's device microphones to improve audio quality for captioners (P7). Others (N=5) worried that setting up captions for all would not be socially feasible: *"I don't see my friends using captioning devices"* (P1). Concern over sightlines persisted for four participants and P9 raised that *"seeing facial expression, seeing if they are angry, upset, happy—you can't get that from captions."*

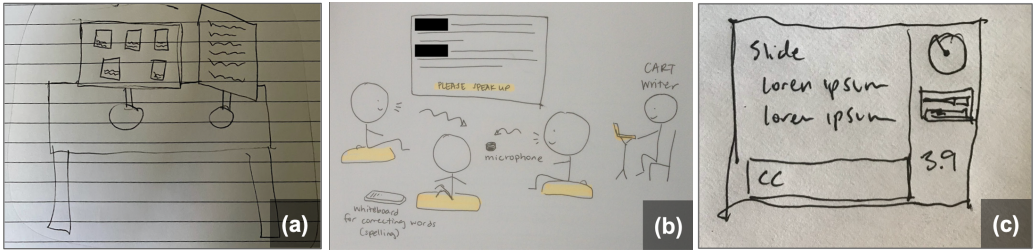

Figure 3: Captioning sketches from of (a) an online captioning setup with captions shown under each speaker and on a separate display (P14), (b) an in-person setup with a shared display and automated feedback for hearing speakers (e.g., "please speak up") (P2), and (c) a customizable captioning interface with speech rate, volume, and lag (P3).

**Online contexts.** While participants had varied reactions to each in-person setup, they overwhelmingly preferred making captions available to all during online conversations. Fourteen participants echoed P14's statement, *"if there's five percent of the captioning that is beneficial to other people, then why not?"* Only one participant (P11) was more hesitant. She explained that if no one else needed captions, she would prefer they were only available to her, *"simply because it's just sort of a personal thing."* Nine participants were excited by the fact that, with online interfaces, all participants could configure captions to meet their personal preferences and access needs.

**Participants' preferred setups.** In addition to the captioning visibility probes we showed, we invited participants to sketch out and reflect on their ideal captioning setups. Certain aspects of space were repeatedly mentioned—four participants stressed the importance of proper lighting and six wanted to be seated at a round table because with *"square, lateral type edges it's harder to look severely to my right or severely to my left. But if it's a more oval shape people are seated in a way that I can see them"* (P13). The form factors of captioning displays were also important to building connection with interlocutors, and participants proposed various novel approaches. P2 considered how to make captioning environments feel cozy (Figure 3b), P7 wished for drone-based captions that hovered over speakers' heads, and three participants imagined the value of captioning glasses, including P9 who pictured hanging out in his living room with friends, *"able to lean back, lean forward, not have to have implements in front of them that makes them focus on one thing instead of looking at everybody."* P11 provided a counterpoint, explaining that her ideal experience is the one she is used to, seated next to a CART writer with the screen *"just between us,"* which she values because *"it's like you are connecting with a person and [...] there is something very human about it."* Participants highlighted how captioning environments and technologies dictate what social interactions will occur and how accessible they can be.

### 4.2.2 Adding Features to Captioning Displays

We discussed six potential features to add to captioning displays: *speech rate, speaker identification, volume, caption lag, overlapping speech,* and *error correction.* For each feature, we introduced and had participants respond to the idea in general before showing a design probe and encouraging the participant to consider a variety of design possibilities. While speaker identification and overlapping speech were somewhat familiar to participants, the other features, as well as considerations as to how participants might like their hearing interlocutors to interact with them, are not captioning standards. We used the probes to explore what additional features participants perceived as potentially useful, to gauge reactions to having hearing interlocutors engage with captions, and to synthesize concrete design takeaways.

**Speech rate (Figure 2a).** Roughly half of the participants (*N*=8) thought speech rate feedback would be valuable for their hearing interlocutors to see, and three participants felt it would also be useful for themselves. P12 suggested that speech rate monitoring *"could be helpful if [a speaker] know[s] that [they] tend to talk really fast and that makes it hard for people to understand [them]"* and P15 saw value in shifting the social burden of telling people to slow down to technology. However, P9 was not optimistic, believing that inflexible social norms lead hearing people to pay attention to *"the content of what they are going to say, not whether they are talking slow or fast."* A key concern raised by six participants was that displaying speech rate could be distracting (our probe included animation) and thus impact comprehension: *"The whole goal is to have things as least distracting as possible in order to maximize the ability to read."* (P14) Future designs could, as three participants suggested, only warn viewers when people speak too quickly or, as P10 proposed, display more caption lines when people speak rapidly so that the reader can catch up.

**Speaker identification (Figure 2b).** Reflecting earlier findings (Section 4.1.3), participants unanimously wanted speaker identification in both automatic and human-generated captions, with seven participants underscoring that speaker identification in their current captioning tools is inadequate. Participants perceived speaker identification as more relevant for themselves than their hearing collaborators, though no one took issue with universal access and five saw it as actively beneficial: *"it would help them be more aware of what it is like when a person cannot hear"* (P5). Ten participants emphasized that color-coding speakers (a feature included in our design probe) is a useful visual shortcut for speaker identification.

**Volume (Figure 2c).** Displaying the current speaker's volume and background noise levels was relatively popular; ten participants were interested, three uninterested and two had mixed feelings. While some participants (*N*=3) wanted to know how loudly they were speaking so that they could self-regulate, more (*N*=7) were interested in providing volume feedback to the group, though four worried the display would be distracting. Three participants hoped that displaying background noise levels might lessen hearing people's tendency to ignore it, lead to a quieter, better setup for captioning—an example of wanting to use the captioning technology to shape social norms and alter the environment. P14 also remarked *"I think hearing people would benefit too because then they would know where the noise is coming from too."* Ten participants independently suggested that they would get more value out of sound identity than volume levels, and desired sound classification integrated with captioning.

**Caption lag (Figure 2d).** DHH participants were interested in conveying how delayed captions are to their hearing interlocutors, though largely did not consider lag to be personally useful. While six participants suggested that seeing the lag would help them make sense of confusing captions, the other nine expressed that they always assumed captions were delayed and therefore did not want feedback. However, eight participants believed that highlighting caption lag for hearing people could support a shared attention to how captions function in practice. P1 stated, *"it's great because they might understand like why I might be jumping in later than I was supposed to,"* and P8 hoped *"this might actually help people put a little bit more buffer time into their speaking."* While our probe conveyed lag in terms of seconds delayed, participants brainstormed other ideas: P4 proposed a *"number of sentences delayed"* metric and P12 imagined a more visual representation: *"It could be like dots indicating every vowel or important recognizable-as-speech sound, [...] something that transforms into the word as the captioning service catches up."*

**Speaker overlap warning (Figure 2e).** Reflecting the fact that captions are not able to capture multiple speakers at once, participants were overwhelmingly interested in an overlapping speaker warning, both for personal (N=14) and group (N=10) use. P12 was the sole participant uninterested in a built-in speaker overlap warning, explaining, *"this feels like more of a social norm thing rather than something that the programming should account for."* Ten participants wanted this information shared with their conversation partners because people *"just get really really excited and start speaking up over each other"* (P2). The other five participants, however, had reservations about the social impact of this technical intervention, with P9 believing *"hearing people want power, so they say, 'Ah, well everybody else can stop talking, I am going to continue'"* and P8 worrying the warnings could have the side effect of shutting down the casual conversations she loves participating in. When considering implementing overlap warnings, participants imagined different roles for this technology: for instance, P15 appreciated that an automated warning could be perceived as less socially disruptive, but P4 proposed adding a blaring siren so that her interlocutors *"all go, 'Oh crap, I need to stop.'"*

**Shared error correction (Figure 2f).** Managing captioning errors has received significant attention in prior work (e.g., [ 6,24,39]), but we sought to explore how participants felt about engaging their direct interlocutors to address errors in real time. Due to time constraints, we only discussed a feature to allow conversation partners to correct inaccurate captions with thirteen participants. Ten participants were interested, and they imagined many benefits of crowd-sourced corrections, such as addressing domain-specific acronyms (P1) or captioning a multilingual workplace: *"[If] someone comes from a similar culture as the speaker, they might be able to input those vocabulary words"* (P2). However, three participants did not think that a group could provide error corrections quickly, and P4 postulated, *"I don't know if you can listen to people speaking, and then also listen to yourself, and also make corrections to captions. I think you need to have someone there dedicated to doing the corrections."* Beginning with our basic mockup, participants brainstormed ways to make error correction useful and readable in real-time, with six people independently suggesting that color could link corrections with their place in the transcript. This process made clear that while participants are interested in shared error correction, it is a complex social and technical problem.

**Ideal interfaces.** Alongside environmental configuration preferences, the sketching exercise we completed with participants highlighted their ideal captioning interface designs. The majority of participants focused on interfaces for online communication, though some considered in-person interface design. Several participants wanted to have access to features we had discussed, such as P1, whose ideal setup (Figure 3c) included speech rate, volume, and lag monitoring, which could be *"individually customizable to display or not, depending on the end user's preferences."* The desire for customization was shared by P8, who posed that it would also be useful to *"choose which features I want and are relevant to me depending on the situation."* Other participants proposed new feature designs: five people independently suggested displaying captions next to each speaker's online video feed (Figure 3a) because it *"eliminate[s] the need to identify the person speaking, if each of them ha[s] their own individual caption"* (P5). P2 wanted to engage her hearing conversation partners in making captioning more effective by using online meeting software outfitted with *"different buttons to say slow down, speak softly, speak up, speak loudly, talk faster, please spell the word."* Participants imagined technical setups that leveraged the unique environment to build new social interactions and feedback for themselves and their interlocutors.

### 4.2.3 Summary and Implications

Participants' responses to our design probes provide considerations for captioning designers and highlight the interrelated factors that shape the utility of captioning tools. When considering how to display captions, participants focused on the tension between shared displays' potential to negatively alter in person conversation dynamics and the isolation and information loss that can come with being the only person accessing captioning. However, this tension largely disappeared when participants considered captioning online conversation, suggesting that videoconferencing is a unique environment which could support socially acceptable, lightweight technical group captioning interventions. Participants were not uniformly excited about all of our design probes. When considering how to best improve DHH peoples' captioning experiences, providing speaker identification and overlap warnings are clear priorities. However, while the majority of participants were not interested in personally using speech rate, volume, or lag feedback, most imagined that they could be useful in guiding hearing people toward more accessible behavior. This, along with interest in shared error correction suggests that technology that shapes group social norms around captioning is worth pursuing. Additionally, participants' responses highlight that preference for caption configuration or new features are dependent on the interactions between social, environmental, and technical factors. Fully accounting for the context that shapes participants' experiences and preferences surrounding captioning can open new avenues for design. For instance, while prior work on visual dispersion has attended to the importance of captioning form factors, [15,34,42,47,52,57], future captioning designers could integrate tenets of DeafSpace [17,79] to consider how to create in a way that matches Deaf environmental and sociocultural norms. Finally, our participants highlight the complex social dynamics these tools could impact, and while many were excited by and interested in trying feedback tools, others remind that not all hearing interlocutors are equally amenable to changing their behavior.

## 5. DISCUSSION

The findings and implications presented above emphasize the social, technical, and environmental factors impacting small group captioning. We have provided an empirical account of DHH participants' experiences and their perspectives on future captioning design. In the discussion below, we synthesize the sociotechnical nature of small group interactive captioning. Further, we reflect on captioning as a group responsibility, the design of future captioning systems, and our study's limitations.

### 5.1 Social, Environmental, and Technical Influences on Small Group Captioning

Throughout our data, participants consistently explained their experiences with small group captioning as shaped by the interaction between social (e.g., DHH people's communication styles, hearing people's mal/adaptive behaviors), environmental (e.g., furniture configurations, features of videoconferencing software), and technical (e.g., delay and accuracy of captions, captioning interface design) factors. Considering all of these factors together provides a more complete understanding of the use and efficacy of captioning technology. For example, our findings show that participants' preferences for captioning form factors are irreducibly determined by the social interactions they permit or prevent and how they shape environments (e.g., needing to have a room with a projector set up as opposed to using a personal laptop to view captions). When considering our participants' experiences of videoconferencing, we found that phenomena such as hearing people correcting captions in the chat can be more completely understood when considering the affordances of online environments that allow for real-time corrections, the social relationships that lead some hearing people to take on caption correction, and the technical failings of captioning that necessitate corrections.

Recognizing that these factors must be considered together to fully contextualize the use of captioning technology has implications for how we as HCI and CSCW researchers work. When formulating research questions, designing studies, analyzing data, and reviewing papers, researchers should consider and seek to account for social, environmental, and technical influences on captioning technology. Many proposed captioning designs have been evaluated out of context (e.g., [7,28]) or in terms of a narrowly defined outcome (e.g., improved comprehension [47] or performance [15]), and future work could complement these analyses with a focus on their social, environmental, and technical contexts. As researchers move to consider the role that hearing people play in conversational accessibility, findings from controlled experiments, such as work done by Seita et al. [65–67], could be contextualized by qualitative work focused on social relationships (e.g., [72]) and the environments in which technology is used (e.g., [37]).

## 5.2 Toward Shared Responsibility for Small Group Captioning Success

Traditionally, researchers have identified DHH people as the primary users of captioning (e.g. [7,15,34]). However, building from our participants' accounts of the impact their hearing interlocutors have on captioning's efficacy, we propose treating captioning as a technology used by *all* members of a group, including hearing people and not solely DHH individuals. Hearing and DHH people both rely on captioning to understand and be understood, but, as participants explained, hearing people often do not recognize their stake in captioning's success. While we believe captioning research should continue to center DHH people—because if captioning does not work, it is DHH people who will lose access—we also seek to reframe captioning as a community-based accommodation [36]. This reframing opens up possibilities for captioning technology designed to support group interdependence [4] by acknowledging that captions cannot work unless people are willing to work *with* them.

This shift provides opportunities to de-center the hearing world norms that are often present in assistive technology design. Many of our participants saw promise for more equitable interactions by introducing hearing collaborators to their world rather than staying in the *"dominant space"* (P8). Participants' proposals to make this shift included simple changes, such as displaying captions for the entire group, and more extreme interventions, including playing loud sirens when hearing people break captioning-friendly norms. Furthermore, participants described the benefits of hearing people learning more accessible communication styles, such as combining typing and ASR captions for casual interactions but aired frustrations around consistently needing to teach these approaches. These sentiments extend Wang and Piper's [72] findings around Deaf/hearing collaboration without accommodations. If rooted in Deaf epistemologies [63], future captioning systems could both teach and reinforce captioning-friendly behaviors to shift labor away from DHH people.

However, we resist embracing captioning for the group without considering potential challenges and opposition. As many of our participants illustrated, there can be high costs to using captioning in an audist world, ranging from social discomfort to workplace barriers, and some remained skeptical that hearing people would ever change inaccessible behaviors. Future work will need to explore the social factors that led some participants to work extensively with hearing interlocutors to collaboratively improve access while others desired captioning solutions that minimized hearing people's involvement. Additionally, some participants were uninterested in changing how they communicate and, as designers of accessible technology, we must respect that technological intervention is not always appropriate or desired. Furthermore, captioning inherently centers spoken conversation. While many of our participants were oral, worked in predominantly hearing workplaces, and socialized with hearing people, some chose to orient their lives around the Deaf community. Lane [45] argues that trying to redirect Deaf people from this Deaf-World is unethical. Regardless of communication mode, however, there are unavoidable interactions in hearing spaces (such as stores and restaurants) where captioning could be a useful tool. Therefore, we must balance building tools to support these interactions without implicitly or explicitly situating oral conversation with hearing people as superior to Deaf-World norms.

## 5.3 Reflections on Future Captioning Design

Building from our call to integrate social, environmental, and technical factors into captioning research and our reframing of captioning as a group technology, we provide concrete design considerations. Specifically, (1) approaches to providing real-time feedback during captioned conversation, (2) opportunities for online communication to advance captioning technology, and (3) discussion of how to design captioning technology for all DHH users.

Our findings suggest that adding real-time feedback and error correction to shared captioning displays are promising areas for future exploration. Participants identified several features that could be especially useful in guiding hearing people's understanding of captioning-friendly behavior (e.g., speech rate, lag), while other features may be more useful in providing DHH people with context on the captioned conversation (e.g., speaker identification, overlapping speaker warnings). How to specifically design and implement each of these features, however, is an open question. A range of options exists such as displaying information directly (e.g., raw decibel levels), integrating information into caption design (e.g., visualizing words that have been spoken but not captioned), or only providing warnings when captioning-friendly norms are breached (e.g., a warning when speakers talk too quickly). Further, there are likely pros and cons to conveying any additional feedback via a shared interface to all conversation participants versus providing individual displays with differentiated feedback—perhaps based on hearing status. Group-generated error correction also merits further study, but many social and technical considerations remain, such as who does the corrections, how can it happen efficiently, and how are corrections integrated into captions. Implementing and testing these features with DHH and hearing users is an important next step to explore questions such as how effective the feedback is at driving behavior change, how to appropriately bring attention to captioning without overwhelming participants, and how receiving feedback impacts all conversation participants' experiences of captioned conversation.

Our study highlights opportunities to evolve captioning tools for online conversations, with a unique capacity to build group-oriented tools. For example, participants perceived that their hearing interlocutors face new constraints online, such as a single audio channel and technical delays, which align with more caption-friendly communication. Online systems could be designed to strengthen these social gains, leveraging the technology-mediated nature of online environments to more easily implement interventions. Other unique online affordances, such as the omnipresence of text chat, the social and technical ease of turning on captions, easily automated speaker identification, and less-settled social norms, could be leveraged to address the distinct disadvantage DHH people face without visual and spatial cues online. Currently many of these features are difficult to implement in person, even with customized hardware such as microphone arrays, and exploring their impact online could help drive priorities in software and hardware development to support captioning users as in-person conversation becomes feasible again.

Finally, future captioning systems should explore ways to allow full participation for all DHH people, regardless of their communication preferences, during captioned conversation. Our study participants who prefer not to voice (8 out of 15), stressed that captions do not support their contributions to a conversation, a concern relevant to the estimated 100,000–500,000 Americans who primarily communicate in ASL [54]. As our findings demonstrate, there are many captioning use cases for people who prefer to sign, particularly as automatic captions become widespread, and it is critical to consider people who do not voice when designing for interactive captioned conversations. This extends beyond simply making it possible to type, as designers must consider how to socially integrate typed contributions into the flow of conversation and account for differences in typing speed as compared to speaking or signing (~50 vs. 160 wpm typing on a touchscreen vs. speaking and signing [3,64]). Future work could explore, for example, allowing people to hold their conversational turn while they type, ways to stream typed contributions as they are generated, and how to help change social expectations around the pace of conversation.

## 5.4 Limitations

Our study has four primary limitations. First, while online recruiting allowed us to expand our geographic reach, we conducted this study during global health, political, and economic unrest, which limited recruitment to those who could spend 90 minutes participating in an online research study. Our 15 participants were all U.S.-based professionals with high-speed internet access. Second, as we explain in Section 3, we chose to conduct this research with DHH participants only, and we do not claim to advance knowledge about hearing people's experiences during captioned conversation. Instead, we explore how DHH people's captioning experiences are impacted by their hearing interlocutors and their preferences for future engagement from hearing people, intentionally giving the power to dictate future design directions to DHH participants only. Future work exploring group captioning tools should involve both DHH and hearing participants. Third, we focused our design probes on contexts where a single DHH person communicates with a group of hearing people, and we do not claim that our findings extend past this scenario. As some participants explained, the conversational dynamic can change when multiple DHH people are in conversation with hearing people. Finally, while DHH people are a large portion of real-time captioning users, they are not the only group that uses captioning as an access tool. We outline findings specific to DHH people, but future work could investigate to what extent these findings are relevant to other captioning users.

## 6. CONCLUSION

In reporting on a formative study with 15 DHH participants, we present an empirical account of DHH people's experiences of captioning during small-group conversation, highlighting the social, environmental, and technical factors that shape the use and usefulness of real-time captioning. Additionally, we outline participant's preferences for the design of future captioning systems, providing design implications regarding captioning as a group technology. Throughout, we discuss participants' experiences of and design preferences for online communication, recognizing it as an environment with unique affordances and considerations for captioning. Our discussion highlights the need to consider social, environmental, and technical context when undertaking captioning research, proposes a shift toward treating captioning as a technology used by groups, and outlines future design considerations. Guided by Deaf and disability studies, we look to a future where DHH and hearing groups use captioning as one of many tools to negotiate conversation accessibility that questions the hearing world's norms.

## ACKNOWLEDGEMENTS

We thank Dhruv Jain for his early feedback and help with piloting, the interpreters and captioners who provided study session communication support, and the University of Washington's Disability Service Office for assistance in hiring captioners and interpreters. This work was supported by the National Science Foundation under Grant No. IIS-1763199, the National Science Foundation Graduate Research Fellowship under Grant No. DGE-1762114, and by the University of Washington CREATE.